\documentclass[10pt]{article}

\usepackage{amsmath}
\usepackage{amssymb}

\usepackage{graphicx}

\usepackage{cite}

\usepackage{color} 

\usepackage[labelfont=bf,labelsep=period,justification=raggedright]{caption}

\makeatletter
\renewcommand{\@biblabel}[1]{\quad#1.}
\makeatother

\date{}

\usepackage{graphicx}

\usepackage{scicite}

\begin{document}

\begin{flushleft}

{\Large
\textbf{Mechanical properties of growing melanocytic nevi and the progression to melanoma}
}
\\ 
Alessandro Taloni$^1$, Alexander A. Alemi$^2$, Emilio Ciusani$^3$, James P. Sethna$^{2}$ Stefano Zapperi$^{1,4}$, Caterina A. M. La
Porta$^{5,*}$\\ 
 \normalsize{{\bf 1} Istituto per l'Energetica e le Interfasi, Consiglio Nazionale delle Ricerche, 
Milan, Italy}\\ 
\normalsize{{\bf 2} Laboratory of Atomic and Solid State Physics, Department of Physics, Cornell University, Ithaca, NY, USA} \\
\normalsize{{\bf 3} Istituto Neurologico Carlo Besta, Milan, Italy} \\
\normalsize{{\bf 4} Institute for Scientific Interchange Foundation, Turin, Italy}\\ 
\normalsize{{\bf 5} Department of Biosciences, University of Milan, Milan, Italy}\\
\normalsize{$^*$ Corresponding author: caterina.laporta@unimi.it}
\end{flushleft}

\section*{Abstract}
Melanocytic nevi are benign proliferations that sometimes turn into malignant melanoma in a way 
that is still unclear from the biochemical and genetic point of view. Diagnostic and prognostic tools are then
mostly based on dermoscopic examination and morphological analysis of histological tissues.
To investigate the role of mechanics and geometry in the morpholgical dynamics of melanocytic nevi, we study a computation model 
for cell proliferation in a layered non-linear elastic tissue. Numerical simulations suggest that the morphology of the nevus is correlated  to the initial location of the proliferating cell starting the growth process and to the mechanical properties of the tissue.  Our results also support that melanocytes are subject to compressive stresses that fluctuate widely in the nevus and depend on the growth stage. Numerical simulations of cells in the epidermis
releasing matrix metalloproteinases display an accelerated invasion of the dermis by destroying the basal membrane. Moreover,
we suggest experimentally that osmotic stress and collagen inhibit growth in primary melanoma cells while the effect
is much weaker in metastatic cells.  Knowing that morphological features of nevi might also reflect geometry and mechanics 
rather than malignancy could be relevant for diagnostic purposes.


\section*{Introduction}

Melanocytic nevi are benign proliferations of melanocytes, the skin cells that produce the pigment melanin. They 
are by definition benign, but 50\% of malignant melanomas arise from pre-existing nevi. Current diagnostic methods
of melanocytic lesions are based on hystopathology and dermoscopy which reveal wide morphological diversity and evolution patterns 
whose origin is still unknown \cite{Baxter2013,Kim2012}.  Leaving aside the controversial cases of Spitz and blue nevi, nevi are categorized from dermoscopic analysis into globular, reticular, structureless brown and mixed patterns (the latter are subdivided into mixed pattern with globular structureless areas in the center or mixed pattern with globuls at the periphery) \cite{Zalaudek2009}. The classical theory of nevi evolution formulated in 1893 by Unna \cite{unna1893} claims that nevi originate from melanocytes proliferating at the  dermo-epidermal junction  (lentigo simplex and junctional nevus) that form nests  (compound nevus) and eventually migrate completely into the dermis (dermal nevus). More recently, Cramer described an opposite model known as the theory of upward migration 
\cite{cramer1988}. Cramer suggested that precursor cells of melanocytes deriving from pluripotent stem cells of the neural crest wander during embryogenesis along nerves into the dermis, mature here and finally migrate as functional melanocytes into the epidermis. Cramer's last migration step, however, was never broadly accepted. More recently Kittler et al. \cite{kittler2000} suggest that nevi can migrate not only vertically but also horizontally, explaining how  nevi can expand in the course of time.  The underlying limitation of these theories is that they are based on hystopathological observation only, and do not reflect the dynamics of an individual nevus. From a diagnostic point of view it would be extremely useful to correlate the morphological features 
of nevi to the degree of malignancy or pre-malignancy. This issue is intricate because nevi proliferate in a complex microenvironment 
that can mediate cell behavior through the composition, structure, and dimensionality of the extracellular matrix (ECM), the polymeric scaffold that 
surrounds cells within tissues. For example, a malignant phenotype can be reverted into a nonmalignant one by specifically blocking aberrant adhesion of the cancer cell to its extracellular scaffold \cite{weaver1997}.   

Recent research shows that mechanical properties of the tumor microenvironment and of the normal tissue can 
influence tumor growth and dynamics, in way that is still poorly understood \cite{helmlinger1997,cheng2009,samuel2011,montel2011,montel2012,Tse2012}. 
Mechanical stresses such as those experienced by cancer cells during the expansion of the tumor against the stromal tissue have been shown to release and activate growth factors involved in the progression of cancer \cite{paszek2005}. Moreover, the stiffness of the matrix surrounding a tumor determines how cancer cells polarize, adhere, contract, and migrate, and thus regulates their invasiveness \cite{hoffman2011}. Forces exerted by cancer cells as they migrate through the ECM has been
quantified accurately using traction force microscopy \cite{zaman2006,koch2012}
Another possibility is that mechanical stresses directly regulate the growth and death rates of cancer cells as shown by Montel et al. who induce osmotic stress by adding dextran, a biocompatible polymer that is not metabolized by cells \cite{montel2011,montel2012}. 
Several studies in the literature indicate important changes in cellular functioning due to osmotic pressure \cite{Racz2007,Nielsen2008}, but the stresses involved  (in the MPa range) were orders of magnitude larger than those (in the kPa range) studied in Refs. \cite{montel2011,montel2012}. 
It is interesting to notice that compressive stresses of slightly less than 1 kPa
applied through a piston were recently found to induce a metastatic phenotype in cancer cells \cite{Tse2012}.
While osmotic pressure may have a different origin than compressive mechanical pressure the effect on cells is exactly the same. This was demonstrated experimentally in Refs. \cite{montel2011,montel2012} by applying osmotic pressure directly on the cells, by adding
dextran to the growth medium, or indirectly, by placing the cells inside a dialysis bag which was then placed into a dextran solution. In both cases, the effect of pressure on the cells was found to be exactly the same. This is not surprising, since osmotic pressure corresponds indeed to a {\it real} mechanical pressure on the membrane due to the collisions with solute molecules.

In a recent paper, Simonsen and coauthors showed that interstitial fluid pressure (IFP) was associated with high geometric resistance to blood flow caused by tumor-line specific vascular abnormalities in xenografted tumors from two human melanoma lines with different angiogenic profiles \cite{simonsen2012}. In another recent paper, Wu and colleagues investigated how nonlinear interactions among the vascular and lymphatic networks and proliferating tumor cells might influence IFP transport of oxygen, and tumor progression \cite{wu2013}. They also investigated the possible consequences of tumor-associated pathologies such as elevated vascular hydraulic conductivities and decreased osmotic pressure differences. All these parameters might affect microenvironmental transport barriers, and the tumor invasive and metastatic potential, opening interesting new therapeutic approaches. In general, understanding the influence of mechanical stress on cancer growth could shed new light on tumor development and progression.

To investigate the role of the environment in the growth of melanocytic nevi, we introduce a computational model of elastic cells proliferating in a layered non-linear elastic medium representing the skin. Several computational models have been introduced in the past to simulate cancer growth in vitro and in vivo. They range from individual cell models \cite{Walker2004,Holcombe2012,galle2005,drasdo2005,drasdo2012}, to lattice models \cite{Simpson2013,Treloar2013,Plank2012} and continuum models \cite{chatelain2011,Plank2012,Eikenberry2009}. The aim of our investigation is to correlate the morphological features of the growing nevus with the location of the originating melanocyte and the mechanical properties of the environment. To this end, a natural strategy is provided by individual cell models where each cell can be mechanically deformed and moves according to mechanical forces due to other cells and to the ECM. Models of this kind have been used in the past to simulate the growth of the epidermis \cite{Adra2010,Thingnes2012} and the growth of cancer cell colonies inside a confining medium \cite{drasdo2012}.

We first report experimental results on melanoma cell lines showing that a low osmotic pressure reduces proliferation, in agreement with other reports for other tumors \cite{montel2011,montel2012}. We have performed experiments on two cell lines, the first obtained from a primary tumor and the second from a metastatic one, both from the same patient. We find that melanoma cells stemming from a primary tumor are less sensitive to pressure than the metastatic ones. 
We find a similar results by growing the same cells in collagen coated plates, simulating the effects of the ECM. 
Melanoma is widely believed to be originating from melanocytes, which are not easily cultivated in vitro.
Therefore melanoma cell lines is the most sensible in vitro model for nevi. We expect that the effect of pressure
on proliferation seen in melanoma cells will be even stronger for melanocytes since melanoma cells are much
more resistant.

Numerical simulations of the model yield different growth patterns for nevi originating in the epidermis, either
close to the basal membrane or to the stratum corneum, or in the dermis. Different initial locations give also rise to different mechanical properties as we show by measuring compressive stresses on the proliferating melanocytes. Compressive stresses are fluctuating strongly inside the nevus, with an average stress that typically increases until the ECM  breaks leading to stress drop. A crucial role in the growth pattern is played by the basal membrane, separating the epidermis from the dermis. Depending on its mechanical properties, the membrane can resist the pressure induced by the nevus or fail leading to the invasion of the dermis. In this case, invasion is not correlated with an increased malignancy of the tumor but just to mechanics. This observation of course does not exclude that malignant phenotypes may be induced by mechanical stresses as previously reported in the literature \cite{Tse2012}. It is in fact known that malignant cells break the basal membrane by matrix metalloproteinases (MMPs). We simulate this process explicitly showing that it accelerates the invasion of the dermis, but the general morphology remains the same.

\section*{Model}

\begin{figure}[ht]
\centering
\includegraphics[width=\textwidth]{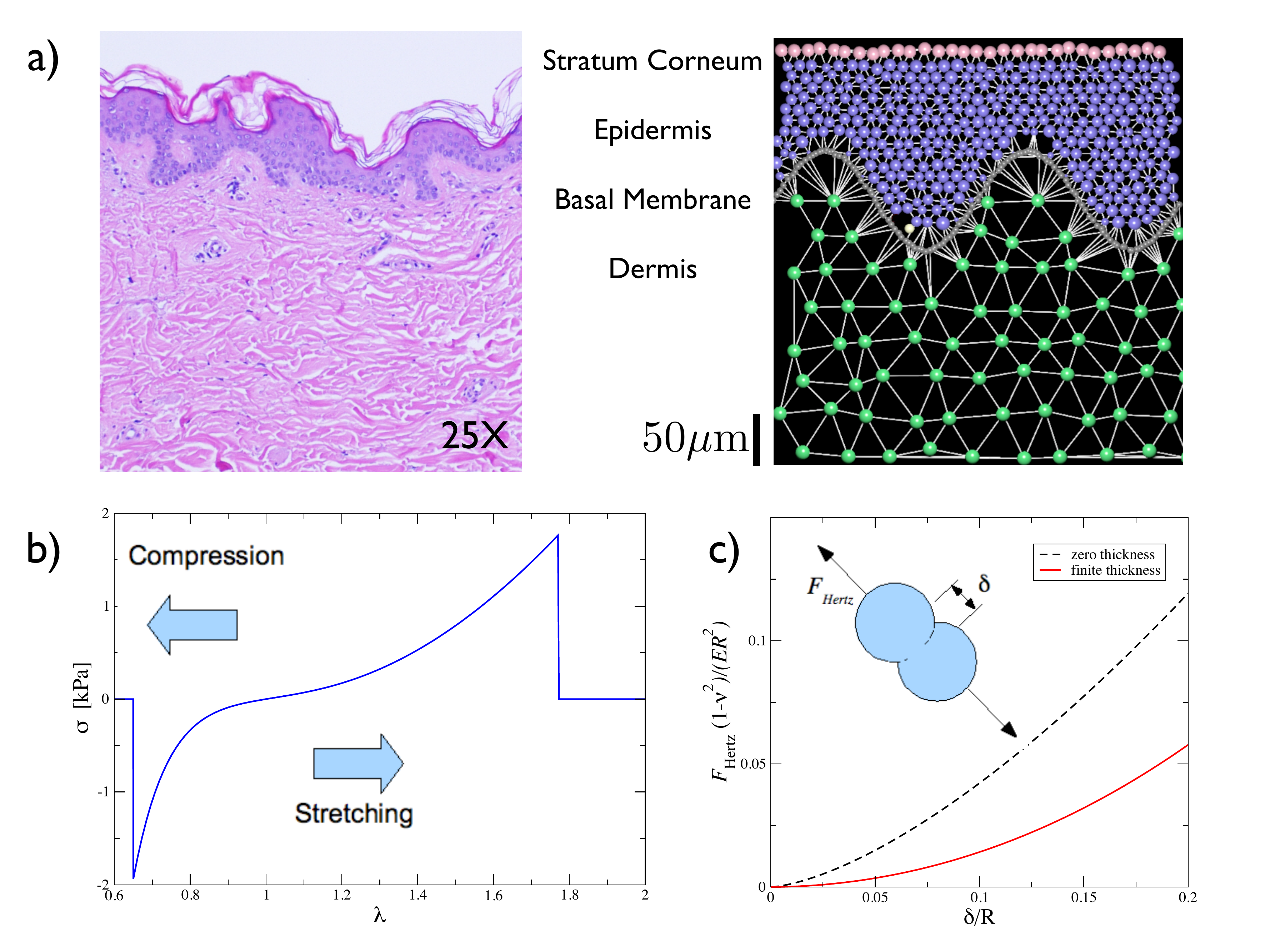}
\caption{{\bf Computational model.} a) A two dimensional section of the skin (left image courtesy of Dr. Claudio Clemente) is simulated as 
a non-linear elastic layered material (right). The top layer is the stratum corneum (pink cells), resting on the epidermis 
(blue cell). The epidermis is separated from the dermis (green cells) by a basal membrane (grey). In a typical simulation we consider a melanocyte (yellow) dupicating inside the skin, either in the epidermis, or in the dermis. b) The connecting fiber have a non-linear elastic behavior with a fracture strength set at $\sigma_Y=1.8 \mathrm{kPa}$. 
c) When in contact, the cells interact by a finite-thickness Hertz law. The corresponding zero-thickness law is
reported for comparison.}
\label{fig:1}
\end{figure}

We simulate an individual cell based model \cite{galle2005,drasdo2005,drasdo2012} for the growth of melanocytic nevi in the skin. We consider a two dimensional vertical section of the skin (see Figure \ref{fig:1}a) consisting of two main layers, the \emph{epidermis} and the \emph{dermis}, separated by the \emph{basal membrane}. The deepest layer of the skin, the hypodermis, is not considered in the present study. In our model, the dermis constitutes the bottom part of the skin, whereas the upper layer is represented by the epidermis and by its most superficial part,  the \emph{stratum corneum}.  The model describes the skin as a highly organized elastic organ, whose mechanical properties are specific of any of its constituent layers.
Periodic boundary conditions are applied to any of the skin layers along the horizontal direction, while open boundary conditions are imposed in the vertical directions.

\subsection*{\underline{Dermis}} The dermis is arbitrarily divided into two anatomical regions: the papillary and reticular dermis. We consider only the papillary layer, whose thickness we set to $150\mu\mathrm{m}$. We model the dermis as a network of cells with an average radius $\sim 15\mu\mathrm{m}$, arranged in a disordered 
lattice and connected by non-linear springs.  The disordered lattice is obtained starting from a triangular lattice, displacing the nodes randomly by a small
amount. The resulting configuration is triangulated by a Voronoi algorithm. Nearest neighbors cells are connected by non-linear springs with a rest length 
equal to the resulting distance between the displaced cells.
Cells here represents fibrolasts, macrophages, and adynocytes present inside the papillary layer, while springs model the effect of the ECM, composed by collagen fibers and elastin. We set the average length of these springs at $\sim 60\mu\mathrm{m}$. The dermis also contains several irregularities such as blood vessels, lymph vessels, nerve endings and skin appendages such as air follicules, small hair muscles and sebaceous glands, but we ignore these details in the present model. The mechanical behaviour of the dermis has been studied experimentally using in-vivo suction and tension tests as reported in ~\cite{Hendriks_2003, Hendriks_phD_2005, Hendriks_2006}. In these studies, the mechanical response of the dermis layer to an external stress was found to obey to the following non-linear \emph{James-Green-Simpson} pressure-displacement relationship \cite{james1975}
\begin{equation}
\sigma = 2C_{10}\left(\lambda-\frac{1}{\lambda^2}\right)+6C_{11}\left(\lambda^2-\lambda-1+\frac{1}{\lambda^2}+\frac{1}{\lambda^3}-\frac{1}{\lambda^4}\right)
\label{stress_NH}
\end{equation}
\noindent where $\sigma$ and $\lambda$ represent respectively the  stress and the stretch of the tissue, the strain being just $\epsilon=\lambda-1$. The expression (\ref{stress_NH}) is derived from the  strain energy density function usually adopted to mimic the non-linear behaviour of the skin ~\cite{James_2000}. Here we use the same law to model the
deformation of the ECM springs, setting their elastic constant to $C_{10}=0.1125 \mathrm{kPa}$ and $C_{11}=0.315 \mathrm{kPa}$, and
introduce a failure stress at $\sigma_Y=1.8 \mathrm{kPa}$, at which the spring breaks. The non-linear elastic behavior of the springs is
illustrated in Fig \ref{fig:1}b, showing that when the stress overcomes $\sigma_Y$ the spring breaks and the stress goes 
to zero. 

\subsection*{\underline{Epidermis}} The epidermis is a multilayered tissue composed of the stratum granulosum, stratum spinosum and stratum corneum. Since the elastic properties of the stratum corneum are cosiderably different from the other two, we will consider it separately. The  thickness of the epidermis (stratum granulosum and stratum spinosum) is  $150\mu \mathrm{m}$ and it is composed mainly by keratinocytes, whose average radius is $\sim 20\mu \mathrm{m}$.  In the model, keratinocytes are placed at the nodes of a disordered lattice, obtained as discussed above for the dermis. The turnover rate of melanocytes is assumed to be much faster than the natural turnover rate of keratinocytes, modelled explicitly in Ref. \cite{Adra2010} but ignored here. This is justified since the typical turnover for
melanoma cells is 1-2 days \cite{baraldi2013} which is much smaller than the turnover of keratinocytes (50-70 days). To simulate the packing of keratinocytes in the epidermis, we chose the radius of each cell after the Voronoi triangulation so that no neighboring cell overlaps. Cells are then coupled by non-linear springs accounting for the neo-Hookean behavior observed in in-vivo experiments  ~\cite{Hendriks_2003, Hendriks_phD_2005, Hendriks_2006}. Experiments suggest that the non-linear pressure-displacement relation in Eq. (\ref{stress_NH}) captures the mechanical properties of the epidermal layer with a good accuracy using the same  parameters values obtained for the dermis ~\cite{Hendriks_2003, Hendriks_phD_2005}.  In the model, we link the cells by non-linear springs with $C_{10}=0.1125 \mathrm{kPa}$, $C_{11}=0.315 \mathrm{kPa}$ and $\sigma_Y=1.8 \mathrm{kPa}$. The average length of the links between cells coincides with the average cell radius, hence two keratinocites may often be in contact and will thus interact as we discuss below. 

\subsection*{\underline{Stratum Corneum}} The stratum corneum is composed of dead corneocytes of an average radius of $\sim 20 \mu \mathrm{m}$. Although it may consists of up to 15-20 layers of corneocytes, in our simplified numerical model only one single layer is taken into consideration.  Experiments ~\cite{Hendriks_2003, Hendriks_phD_2005, Hendriks_2006} have provided evidence that the stratum corneum mechanical properties are consistent with a linear Hookean law.
Here, we consider elastic cells linked by linear springs with Young's modulus $E=1.5 \mathrm{kPa}$  and failure stress set to $\sigma_Y=2 \mathrm{kPa}$. The stratum corneum cells are linked to the epidermis by linear springs whose Young's modulus is $E=1.6125 \mathrm{kPa}$. 

\subsection*{\underline{Basal membrane}} The basal membrane constitutes and elastic sheet separating the dermis from the epidermis. It has a complex structure composed of the lamina reticula and lamina basale connected by collagen fibrils; usually the thickness is around $0.5-1\mu \mathrm{m}$. In our model the basal membrane is represented by an assembly of fictitious cells of radius $0.5\mu \mathrm{m}$, connected by linear springs with Young's modulus which we set either $0.15 \mathrm{kPa}$ or $0.3 \mathrm{kPa}$. The failure stress is of the springs is set to $\sigma_Y=2 \mathrm{kPa}$. The fictitious cells composing the basal membrane have Hertzian repulsive interactions with the other cells thus providing an effective barrier to their motion. To represent the highly corrugated structure of the membrane, we use a simple periodic function with amplitude equal to $7.5\mu\mathrm{m}$ and period $30\mu\mathrm{m}$. Dermis and epidermis cells are connected to the basal membrane from below and above respectively, these springs obey to a linear Hookean force with Young's modulus equal to $E=0.18725 \mathrm{kPa}$. In the model, the basal membrane can be broken mechanically, 
but we also consider the production of MMPs. To model this effect, we stipulate that at each time step each cell composing the nevus can dissolve one link with
a probability that decreases with the distance from the cell:
\begin{equation}
P_{MMP}=p_{break}\left(\frac{r_{break}}{r}\right)^2,
\label{MPP_prob}
\end{equation}
with $p_{break}=10^{-2}$ and $r_{break}=30\mu \mathrm{m}$.

\subsection*{\underline{Cell mechanics}}
Cells are modelled as non-linear elastic spheres and, for the sake of simplicity, we employ the same mechanical constants
for all the cells composing the epidermis the dermis and the nevus. When two cells are in contact they repel
by a finite-thickness Hertzian law ~\cite{Long_2011}
\begin{equation}
F_{Hertz} =\frac{4}{3}\frac{E}{(1-\nu^2)}\sqrt{R}\delta^{3/2}\frac{\left(1+1.15\omega^{1/3}+9.5\omega+9.288 \omega^2\right)}{1+2.3\omega},
\label{Hertz}
\end{equation}
\noindent where $E$ is the cell's Young's modulus, set to $E\simeq 1 \mathrm{kPa}$ \cite{lekka1999},  $\nu$ is the Poisson's ratio, $R\equiv \frac{R_1R_2}{R_1+R_2}$ with $R_1$ and $R_2$ radii of the contact cells, $\delta$ is the indentation depth and $\omega=0.1\frac{\delta}{R}$ (see Fig. \ref{fig:1}c). We notice that cells are often modelled as incompressible ($\nu=0.5$)
\cite{lekka1999}, but the value of Poisson ratio is difficult to estimate experimentally. Individual cell models use compressible
cells as we do and employ $\nu<0.5$ \cite{drasdo1995,drasdo2012}. Here we use $\nu=0.33$ but its precise value is not really 
important in the framework of our simulations: changing $\nu$ would just imply a modification 
of the pre-factor of Eq. \ref{Hertz} , leaving qualitatively unaffected the numerical results hereby displayed. We notice that, although Eq.(\ref{Hertz}) refers to the experimental situation of a  microsphere indenting a finite-sized soft gel, it captures the non-linear corrections to the usual Hertzian law, needed when large deformations occur.  In most simulations, nevi cells are interacting by Hertzian repulsion only, while the other cells are typically linked also by non-linear spring. We have also tested for the effect of adhesion between nevi cells introducing a radial force $f_{adhesion}= k_{ad}\delta$,  with $k_{ad}=1.5\times 10^{-5}\mathrm{N/m}$.  Adhesive forces only act when the indentation depth $\delta$ is larger than $-20\mu\mathrm{m}$.

\subsection*{\underline{Interactions with the ECM}}

In the simpler implementation of the model, nevi cells have no direct interaction with the ECM. An indirect interaction
exist since nevi cell interact with other cells which are held together by the ECM. This indirect interaction is, however, 
weak inside the dermis. To overcome this limitation, we also consider a direct interaction between nevi cells and the
ECM. In practice, we follow the same strategy used to model the basal membrane and place three fictitious cells with
radius $5\mu \mathrm{m}$ on each ECM link. In this way, nevi cells interact directly with the ECM.

\subsection*{\underline{Growth dynamics}}
The quasi-static growth of the nevus is simulated by randomly selecting a single cell among those belonging to the dermis or epidermis.  This cell represents the first proliferating cell and therefore does not present any linkage with the surrounding skin network. The growth process is obtained by duplication of the initial cell, so that the new cell is still in contact with the first one, but its growth direction is picked at random, i.e. lattice-free: such a choice is motivated by the fact a proliferative population will always eventually fill the lattice, suppressing  cell-to-cell crowding effects \cite{Plank2012}. The insertion of a new cell encompasses an overall rearrangement of the entire network, which is achieved by minimization of the stress within the  system.  Since growing cells are completely disconnected from the surrounding skin tissue, they only interact through Hertzian-like forces (Eq. \ref{Hertz}) and possibly through adhesion. Two different choices of protocol are applied to the duplication mechanism: a cell can either be picked randomly among those composing the nevus, or the cell can be selected with a rate that depends on the compressive stress. In particular,  the probability $P_{i}(N)\propto e^{-\frac{\sigma_i(N)}{\langle \sigma(N)\rangle}}$  is assigned to each  cell, where $\sigma_i(N)$ represents compressive stress of the $i$-th cell and $\langle \sigma(N)\rangle$ is the average stress. Finally, the cell that duplicates is selected  with probability  $P_{i}(N)$.

\section*{Results}

\subsection*{\underline{Experiments}}
\subsubsection*{\it Osmotic pressure affects the growth of primary melanoma cells}
\begin{figure}[ht]
\centering
\includegraphics[width=\textwidth]{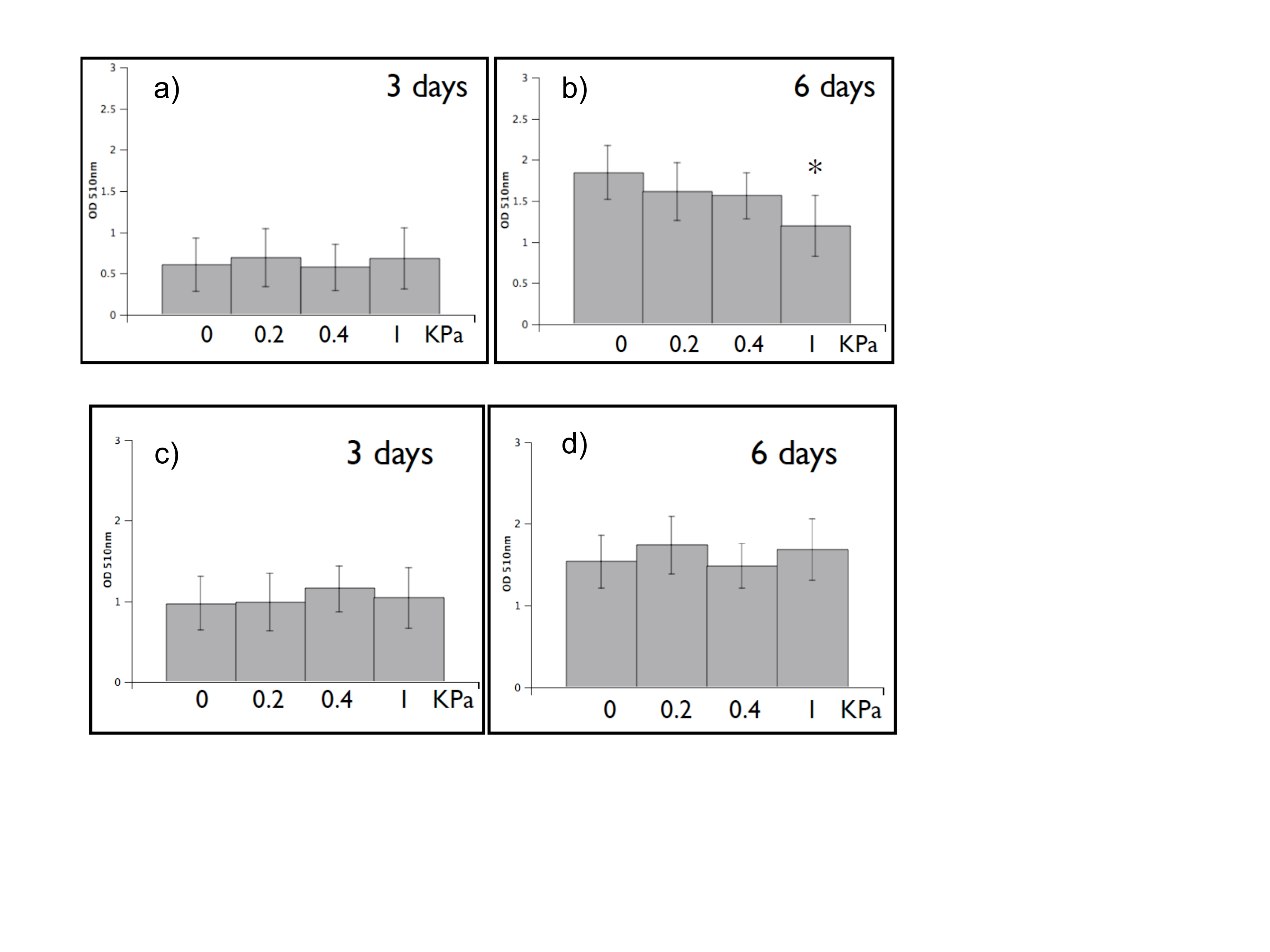}

\caption{{\bf Effect of osmotic pressure on cell proliferation in melanoma.} 500 cells/well were plated on 96 multiwells (IgR39 and IgR37). The day after plating the cells were submitted to different osmotic pressure (from 0.2 to 1kPa) for 3 or 6 days. At the end of the incubation the cells were fixed with 50\%TCA for 2hours at 4C and air-dry at room temperature. Thus, the cells were incubated with 0.05\% SBR solution for 30 minutes at room temperature and then quickly rinsed four times with 1\% acidic acid to remove unbound dye. Finally, the protein-bound dye was solubized with 10mM TRIS and OD was measured at 510nm with microplate reader 550 (Bio-Rad). a) The growth of IgR39 (non-metastatic) cells is not affected by pressure after 3 days, b) but cells grow significantly less after 6 days. c) IgR37 (metastatic) cells are unaffected by pressure both after c) 3 or d) 6 days. 
Statistically significant results according to the KS test ($p<10^{-2}$) are denoted with $*$.}
\label{fig:2}
\end{figure}

We test the effect of low osmotic pressure (from 0.1kPa to 1kPa) on cellular proliferation of primary and metastatic human melanoma cell lines, IgR39 and IgR37, respectively at short (3 days) and long time (6d ays). As shown in Fig. \ref{fig:2}, the treatment with 1kPa for 6 days decreases the cellular proliferation of IgR39
while IgR37 are only slightly affected.  According to Table1, the osmotic pressure increases cell death to necrosis while the percentage of cells in early apoptosis slightly decreases at low pressures and then returns to the 
original value at higher pressure (Table1). We are tempted to attribute this last result to a statistical fluctuation.

\begin{figure}[ht]
\centering
\includegraphics[width=\textwidth]{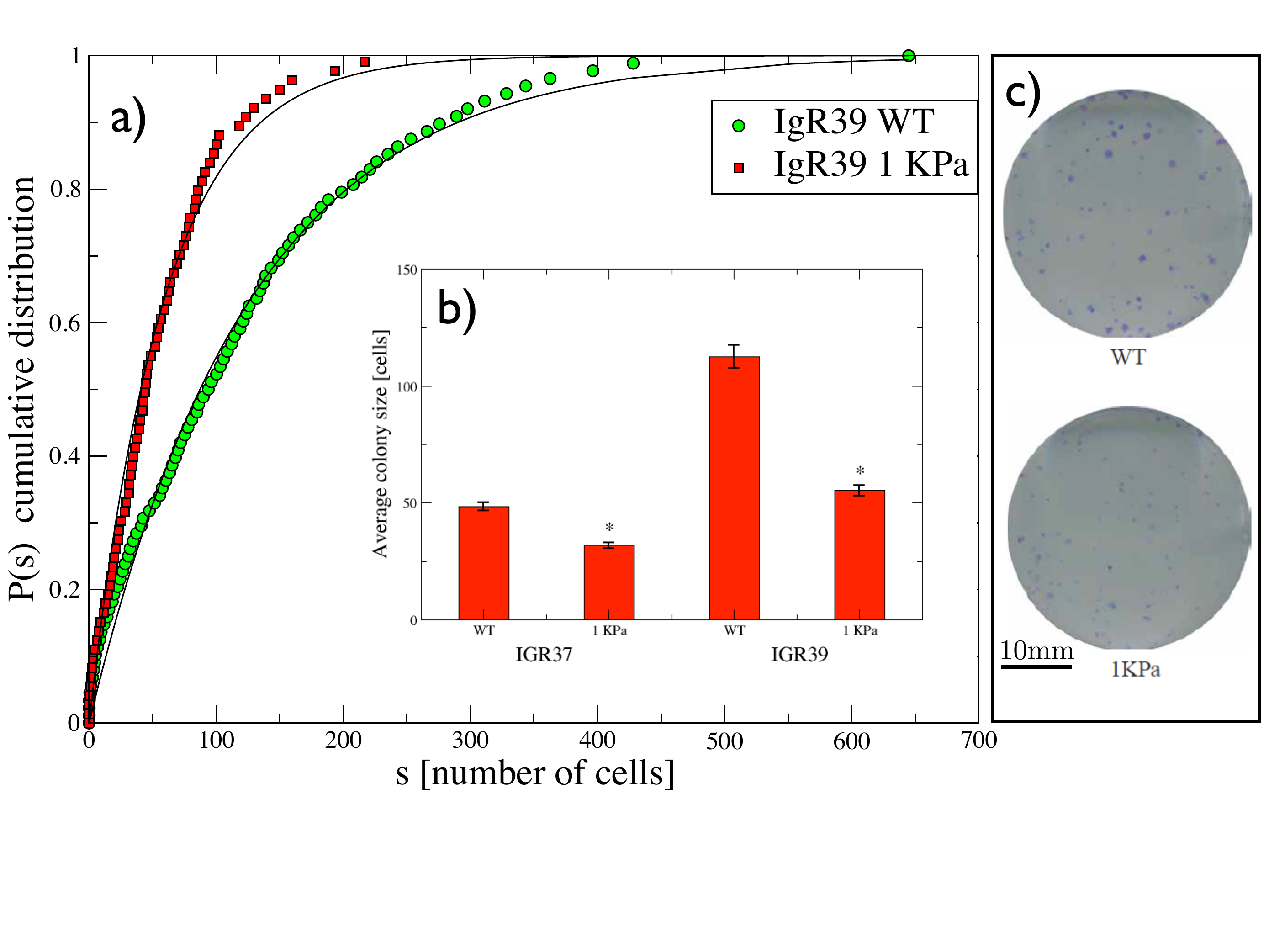}

\caption{{\bf Effect of osmotic pressure on colony formation in melanoma.} a) The cumulative distributions of colony size obtained from IgR39 cells under 1 kPa osmotic pressure with respect to the control (0 kPa). The curves are the fit with a continuous time branching process model (see Ref \cite{baraldi2013}) yielding a division of rate of $\gamma=0.60$ that is reduced to $\gamma=0.51$ under osmotic pressure. b) The average value of
the colony size distribution with the associated standard error for IgR39 and IgR37 cells. 
Statistically significant results according to the KS test ($p<10^{-2}$) are denoted with $*$.
c) The images show two representative examples of the colonies for 0 kPa and 1kPa conditions.}
\label{fig:2bis}
\end{figure}

We also perform a colony formation experiment using the crystal violet assay on primary and 
metastatic melanoma cells, following the prescriptions of Ref.\cite{baraldi2013}. In Fig \ref{fig:2bis}a we report 
the cumulative colony size distribution $P(s)$, where $s$ is the number of  cells in each colony, for IgR39 cells with or without osmotic pressure.  Each distribution is obtained by combining the result of colonies obtained in six different wells. Typical images of the colonies are reported in Fig \ref{fig:2bis}c. We fit the distributions with the solution of a continuum time branching process in which each cell divides with rate $\gamma$  which yields a cumulative
colony size distribution after $t$ days given by \cite{baraldi2013}
\begin{equation}
P(s) =(1-e^{-\gamma t})^{s} ~~\mbox{ for } s \geq 1.
\end{equation}
From the fit, we estimate the average division rate as $\gamma=0.60$ divisions/day in normal conditions and $\gamma=0.51$ divisions/day under an osmotic pressure of 1kPa. In Fig. \ref{fig:2bis}b
we report the average colony size with its standard error for IgR39 and IgR37 cells with and without osmotic
pressure. While osmotic pressure hinders colony growth in both cases, the effect is much stronger for primary
tumor cells than in metastatic ones.

\subsubsection*{\it Effect of collagen on the growth of melanoma cells}

\begin{figure}[ht]
\centering
\includegraphics[width=\textwidth]{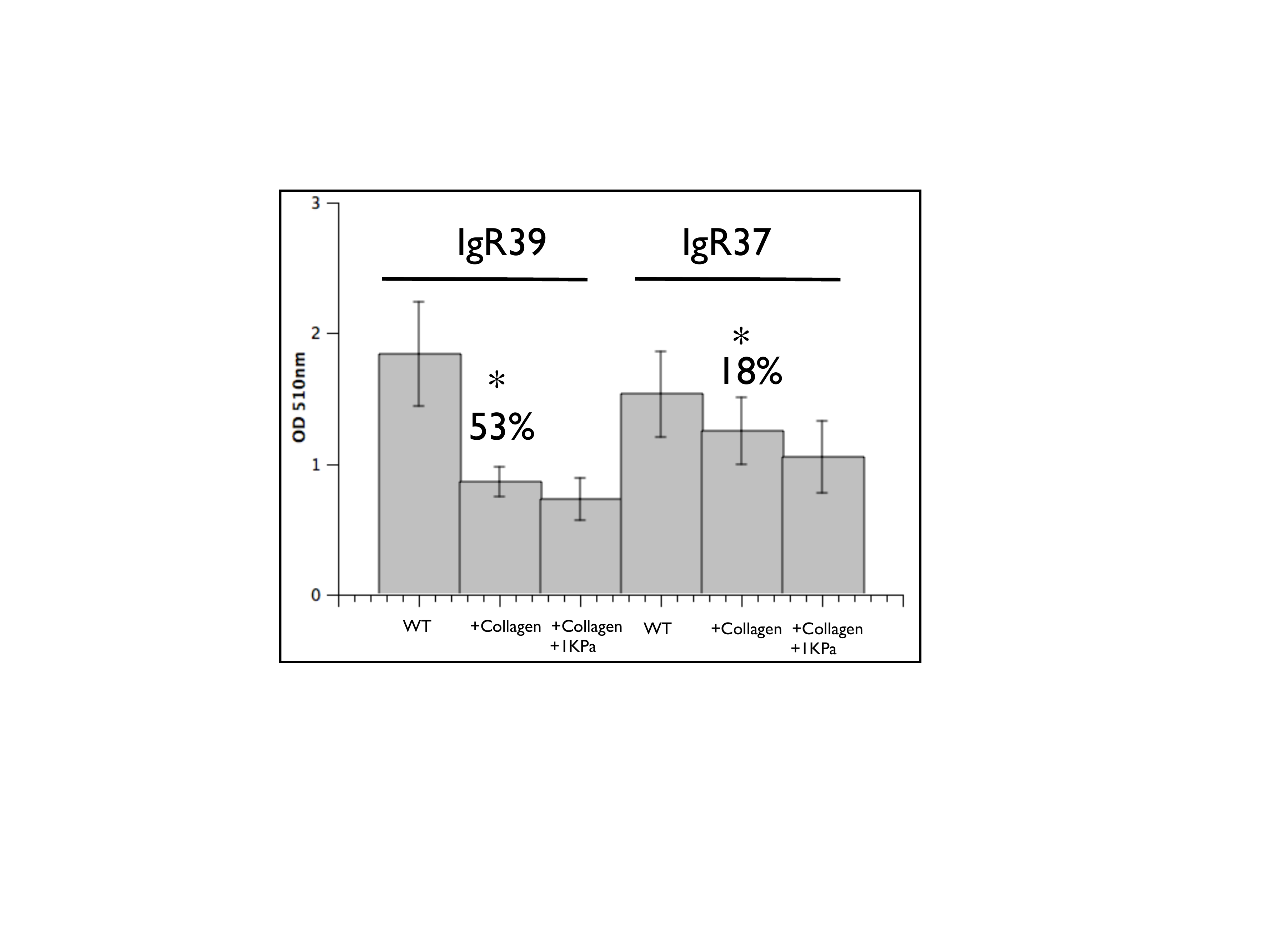}

\caption{{\bf Effect of collagen coating on cell proliferation in melanoma.} 
500 cells/well were plated on 96 collagen coated multiwells (IgR39 and IgR37) and submitted to 1kPa osmotic pressure for 6 days.  Cell growth was evaluated as in Fig.\protect\ref{fig:2}. The graph shows that 
the presence of collagen slows cell growth both in primary (IgR39) and metastatic (IgR37) cells but the
effect is much stronger for primary cells. Additional osmotic pressure does not significantly alter the
results. Statistically significant results, according to the KS test ($p<10^{-2}$), are denoted with $*$.
}
\label{fig:2tris}
\end{figure}

To confirm that the general effect of stress we observe on growing melanoma cells is not restricted to osmotic pressure, we perform proliferation experiments using collagen coated plates simulating the effect of to the extracellular matrix on the cells. As shown in Fig. \ref{fig:2tris}, the presence of collagen significantly reduces
proliferation in primary melanoma cells (IgR39) while the effect is much weaker in metastatic cells (IgR37). 
We next induce an osmotic pressure of 1kPa in the collagen coated plates and find no additional significant
result with respect to the case with collagen alone. We can interpret this result by assuming that collagen already induces
a strong stress on the cells so that additional osmotic stress has no significant effect of the growth. We can thus suggest that osmotic pressure and collagen 
both act in similar ways on melanoma cells: they reduce proliferation but the effect is much stronger in primary
cells than in metastatic ones.

\subsection*{\underline{Numerical simulations}} 

\subsubsection*{\it Morphology of melanocytic nevi depends on the location of the initiating cell}
\begin{figure}[ht]
\centering
\includegraphics[width=\textwidth]{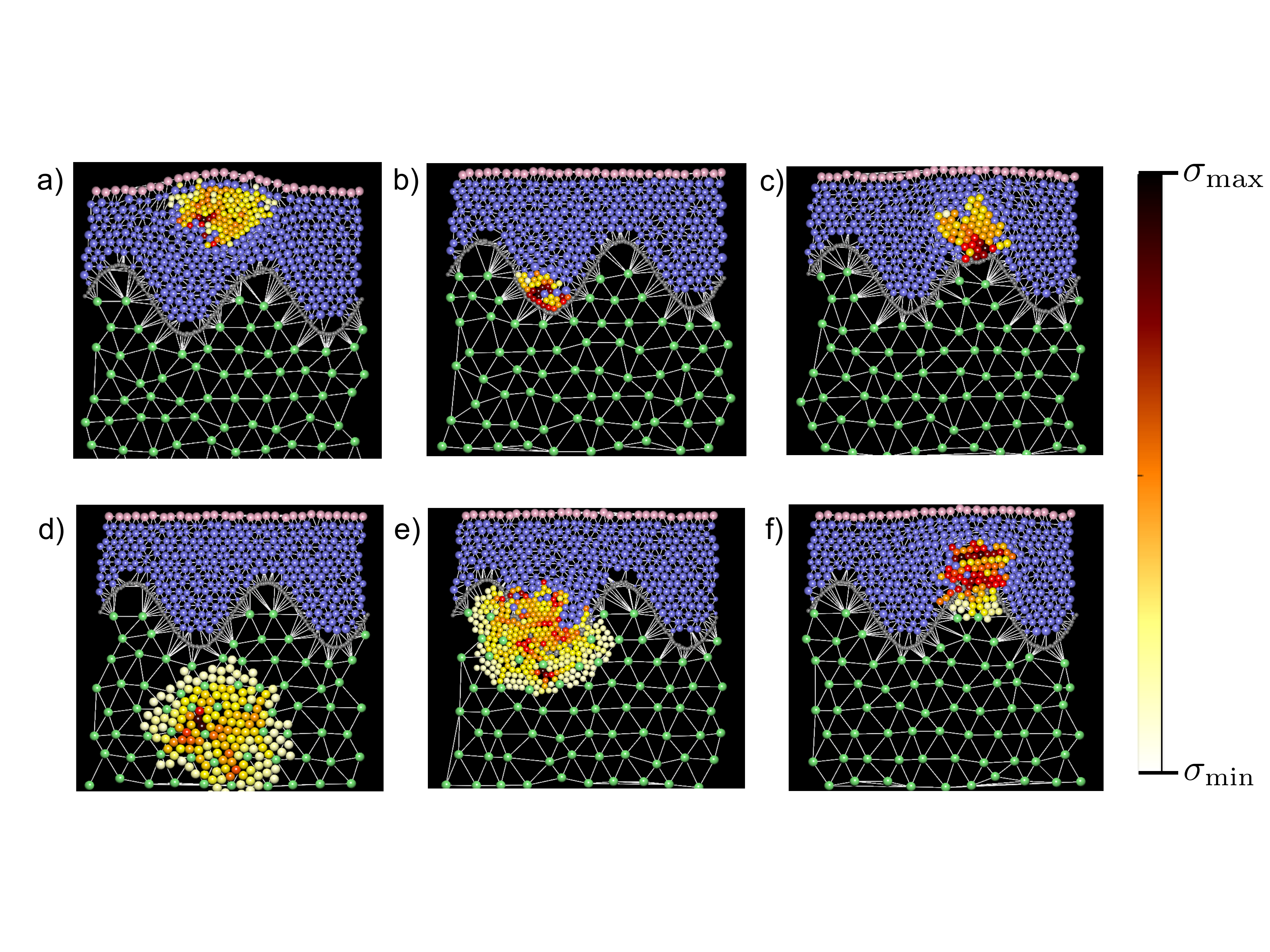}

\caption{{\bf Morphology of nevi for different locations of the initiating cell for random growth.} Illustration of the results of numerical simulations for nevi grown 
from melanocytes located in different positions in the skin and for different mechanical properties of the basal membrane. Here growth occurs
randomly and independently on compressive stress. Growing melanocytes are shown with a varying color that reflects their compressive stresses $\sigma$ according to the color bar. The maximum $\sigma_{\rm max}$ and
minimum  $\sigma_{\rm min}$ for each configuration are equal to (in kPa):
a)$0.61-0.77$, b)$0.0043-0.21$, c)$0.085-0.57$, d)$0.0013-0.149$, e)$0.0047-0.37$, f)$0.0067-0.72$.
a) Nevi grown from  melanocytes residing inside the epidermis  tend to grow horizontally and do not spread towards the basal membrane.
b) Nevi grown on the minima of a strong basal membrane  tend to grow roughly parallel to the membrane itself.  c) Nevi growing from the maxima of a strong basal membrane tend to grow vertically in the epidermis. d) Dermal nevi tend to have radial shape. When the basal membrane is weak, e) nevi growing from the minima of the basal membrane invade the dermis in a radial fashion, while f) when they start from maxima of the basal membrane the invasion of the dermis occurs more vertically. 
}
\label{fig:3}
\end{figure}

We perform numerical simulations of the growth of melanocytic nevi for different conditions. In particular, we study the
morphology resulting from the growth using different locations of the initial seed. As illustrated in Fig \ref{fig:3}, we 
consider four type of initial locations: deep inside the epidermis (Fig. \ref{fig:3}a and movie S1), in the epidermis
but close to the basal membrane, either in the minima (Fig. \ref{fig:3}b, Fig. \ref{fig:3}e and  and movies S2-S3) or on
the maxima of the membrane  (Fig. \ref{fig:3}c, Fig. \ref{fig:3}f and  and movie S4-S5), or inside the dermis (Fig. \ref{fig:3}d  and movie S6).
For initiating cells close to the basal membrane, we consider two different scenarios with a strong (Fig. \ref{fig:3}b and \ref{fig:3}c) or
weak membrane (Fig. \ref{fig:3}e and \ref{fig:3}f). Fig. \ref{fig:3} indicates that the morphology of the nevus depends substantially
on the location of the initiating cell. When the nevus starts up in the epidermis the growth is mostly horizontal and far from
the dermis (Fig. \ref{fig:3}a). For initiating cells residing close to a strong basal membrane the growth follows the profile
of the membrane (Fig. \ref{fig:3}b) or expands vertically in the epidermis (Fig. \ref{fig:3}c) depending on whether the initiating
cells lies in the minima or maxima defined by the membrane profile. When the membrane is weak, however, it is easily broken by
the growing cells that then invade the dermis (Fig. \ref{fig:3}e and Fig. \ref{fig:3}f). The growth is more radial when the initiating
cells is in the minimum of the basal membrane (Fig. \ref{fig:3}e) and more vertical when it lies close to the maximum (Fig. \ref{fig:3}e).

\subsubsection*{\it Mechanical stresses in growing melanocytic nevi}
\begin{figure}[ht]
\centering
\includegraphics[width=\textwidth]{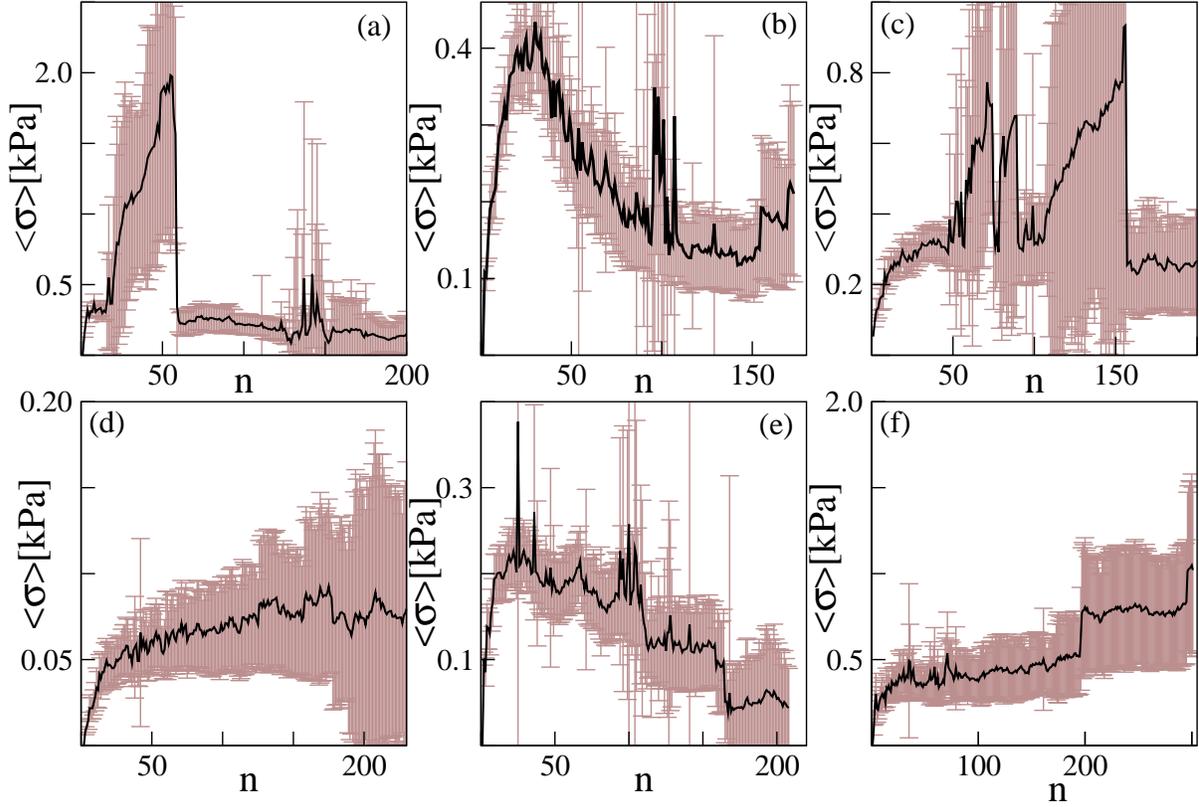}

\caption{{\bf Mechanical stresses in randomly growing nevi.} We report the evolution of the average compressive stress
experienced by the melanocytes composing the nevus as a function of the size of the nevus, quantified by 
the total number of cells $n$, for the same conditions as in Fig. \protect\ref{fig:3}. The error bars represent the standard error of the mean. The stress is averaged over all the cells in a nevus and over at least ten 
statistically independent realizations of the growth process.
The different panels represent different initial locations: a) in the middle of the epidermis, b) in the minima of
a strong basal membrane, c) in the maxima of a strong basal membrane, d) in the dermis, e) in the minima of
a weak basal membrane, f) in the maxima of a weak basal membrane.  }
\label{fig:4}
\end{figure}

We record the compressive stresses sustained by the melanocytes during the growth of the simulated nevi. A pictorial view of the stress in each cell is shown in Fig. \ref{fig:3}, where the color scale depicts compressed cells with a scale going from yellow (low stress) to red (high stress). The images show that stresses are in general heterogeneously distributed and,  as illustrated
by the supplementary movies, fluctuate considerably after each division step. We notice that typically the most stressed cells are located in the
inner part of the nevus, reproducing a general feature observed in tumor spheroids \cite{montel2011,montel2012}. Furthermore, we see 
the formation of stress chains, typical of granular assemblies \cite{howell1999}.
To better quantify the evolution of the compressive  stress we measure the average compressive stress $\sigma$ experienced by the cells as a function of the number of cells in the nevus. The results reported in Fig. \ref{fig:4}  correspond to the different cases explored in Fig. \ref{fig:3}. In all cases, pressure builds up as cells are duplicating until at some point the connective tissue breaks drastically reducing the average compressive stress experienced by the cells. This effect is particularly clear in the case of a nevus growing in the epidermis (see Fig. \ref{fig:4}a). The large error bars reported in Fig. \ref{fig:4} indicate that stresses vary widely between individual cells but also among different realizations of the process. 

\begin{figure}[ht]
\centering
\includegraphics[width=\textwidth]{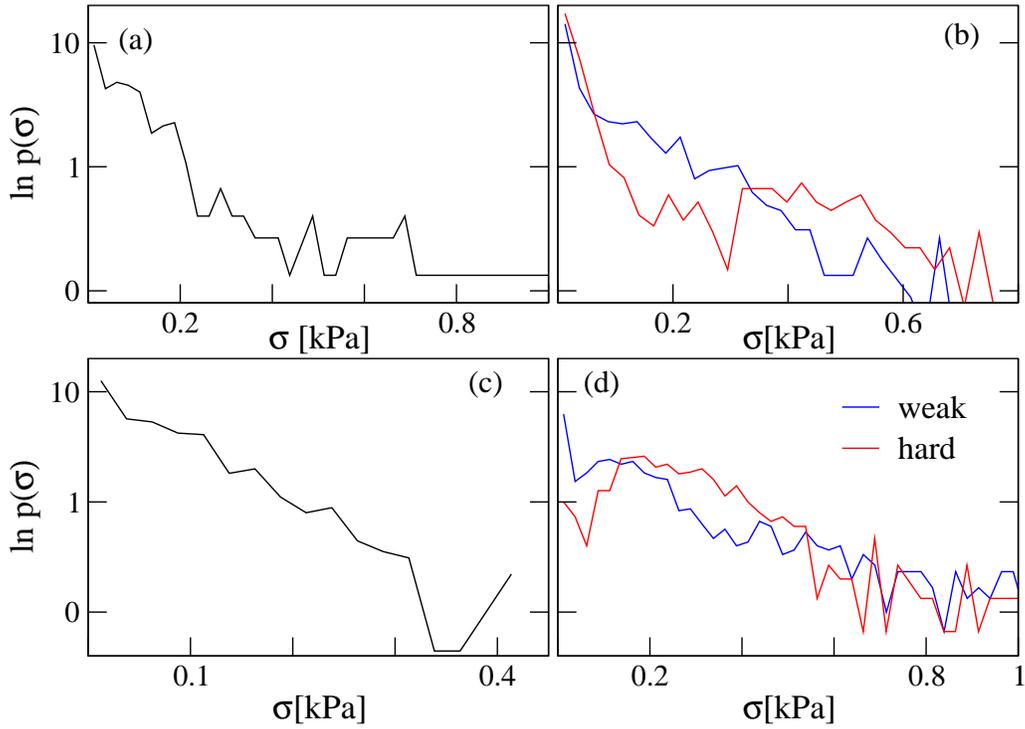}

\caption{{\bf Compressive stress distribution in nevi.} We report the distribution $\rho(\sigma)$ of the compressive stresses
$\sigma$ (in kPa) experienced by melanocytes in a nevus. The results are sampled over different realizations of the process. 
Since the distribution also changes with time, we only consider the time at which the average compressive stress $\langle\sigma\rangle$ is highest.  The different panels represent different initial conditions corresponding to Fig. \protect\ref{fig:3}: a) in the middle of the epidermis, b) in the minima of the basal membrane (for strong and weak membranes) c) in the dermis, d) in the maxima of
the basal membrane (for strong and weak membranes).  }
\label{fig:5}
\end{figure}

To better quantify the stress fluctuations experienced by individual cells, we report
in Fig. \ref{fig:5} an estimate of the distribution (probability density function) $\rho(\sigma)$ of the compressive stresses 
$\sigma$ for different initial conditions. The distribution is sampled over different realizations and over different cells.
Since the distribution is expected to change with the number of cells present in the nevus (as shown from the error bars
in Fig. \ref{fig:4}), we consider only the case for which the average pressure is highest (the peak stress in each curve
in Fig. \ref{fig:4}). In all cases the distribution displays an approximately exponential tail (i.e. $\rho(\sigma) \simeq \exp(-\sigma/\sigma_0)$).

\begin{figure}[ht]
\centering
\includegraphics[width=\textwidth]{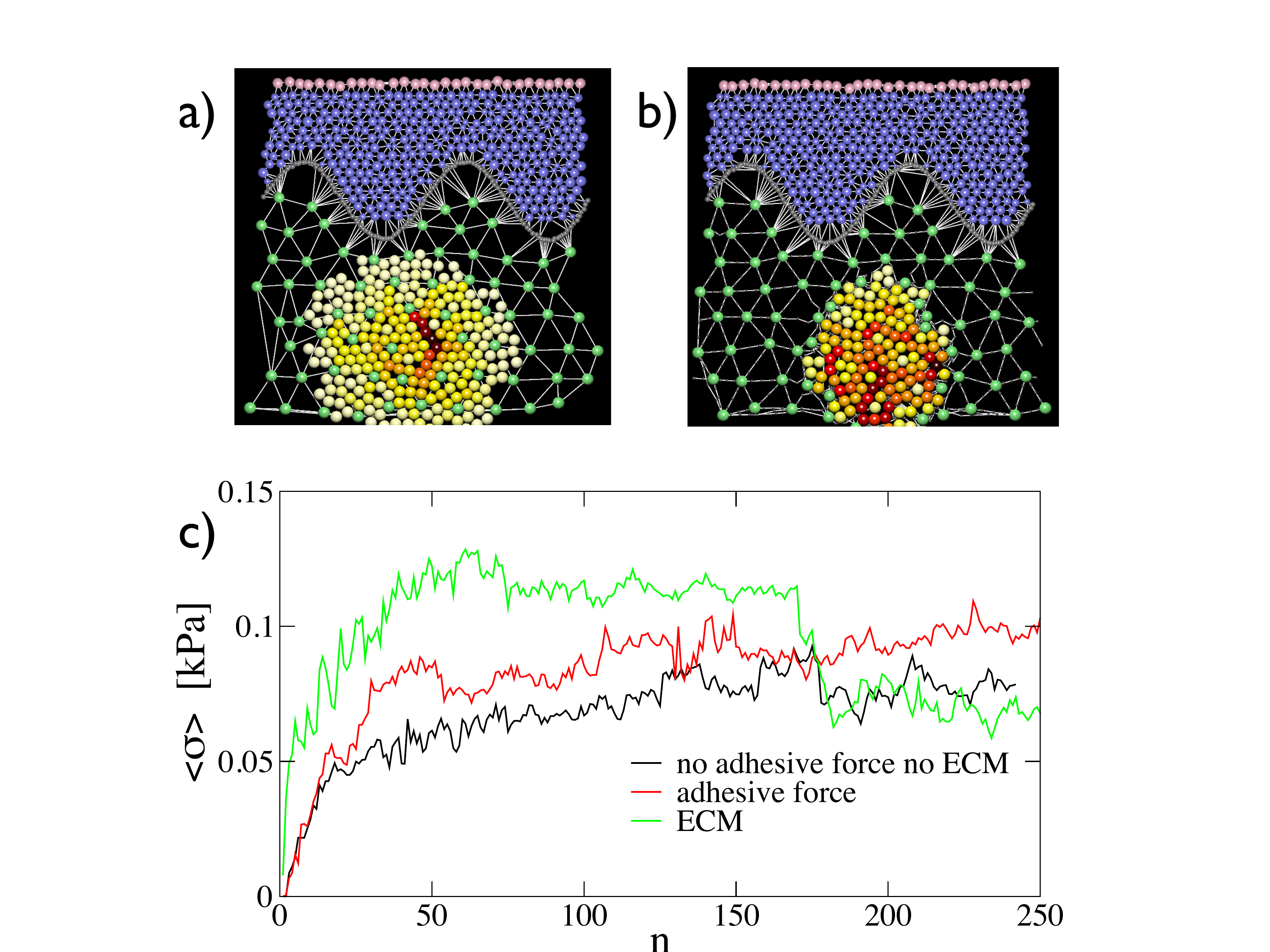}

\caption{{\bf Role of adhesion forces and direct interactions with the ECM.} Morphology of a nevus grown in the dermis a)
in presence of adhesive forces between nevi cells and b)  considering direct interactions with the ECM modelled as discussed in
the methods section. c) The evolution of the average compressive stress experienced by the melanocytes composing the nevus for cases a) and b) is compared with a simulation in which no adhesive forces and  interactions with the ECM are present.}
\label{fig:ecm}
\end{figure}

Compressive stresses are much higher for nevi growing in the epidermis than in the dermis. This is due to the general assumption made in the model that cells are more packed in the epidermis than in the dermis. When we include a direct interaction between the nevi cells and the ECM, as discussed in the model section, compressive stresses in the dermis increases but the general features of the results are unchanged (see Fig. \ref{fig:ecm} and Movie S7). We have also tested for the effect of adhesion between
nevi cells and found very little differences with respect to the case in which adhesion is absent (see Fig. \ref{fig:ecm}).

\subsubsection*{\it Rupture of the basal membrane} 
\begin{figure}[ht]
\centering
\includegraphics[width=\textwidth]{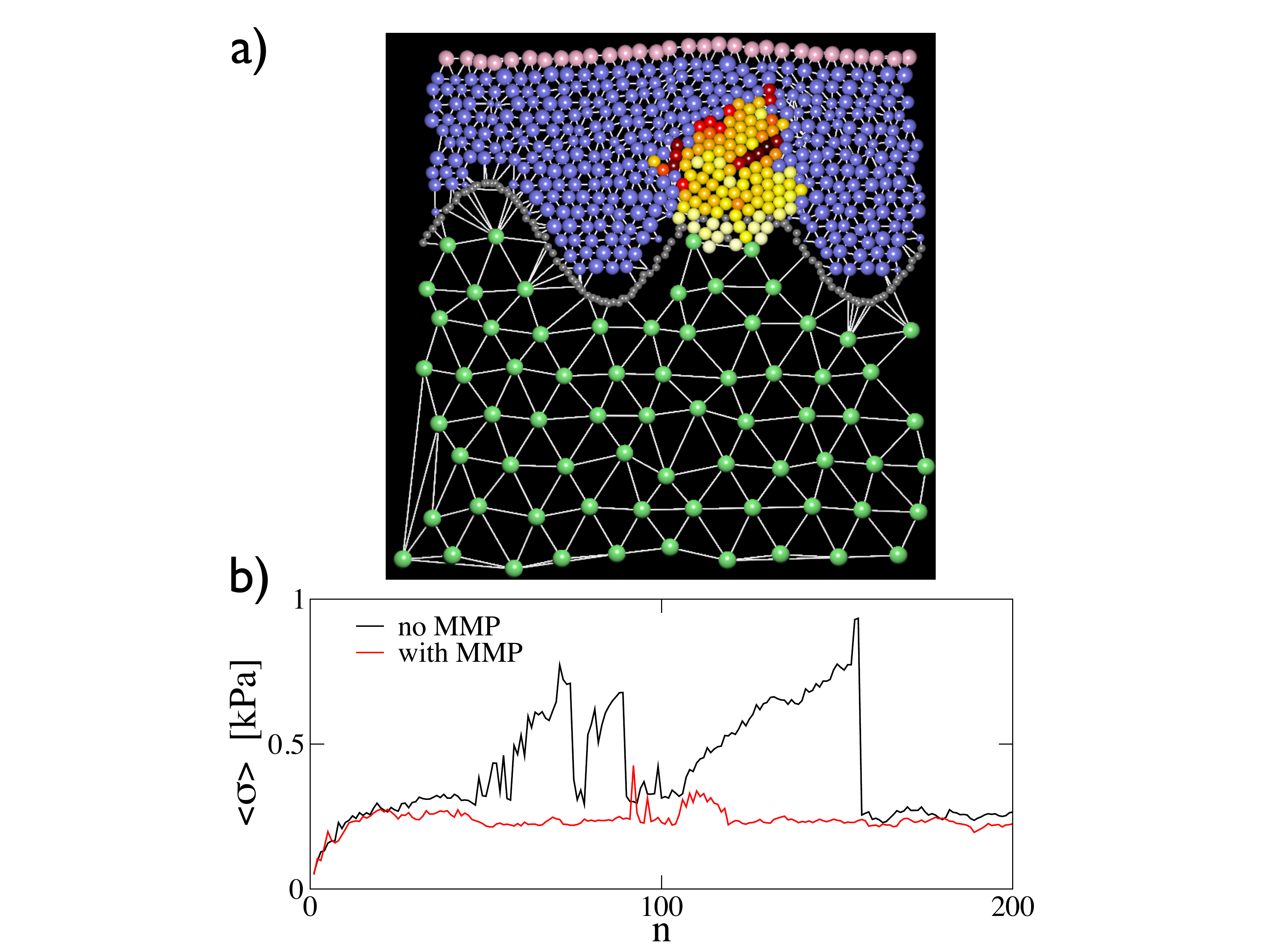}

\caption{{\bf Model of MMP induced breaking of the basal membrane.} a) We report the typical morphology of a nevus starting
from the epidermis with a strong basal membrane that can however be broken by MMPs. b) The evolution of the average compressive
stress in presence of MMPs is compared with the case in which MMPs are not present. }
\label{fig:mpp}
\end{figure}

The results reported in Fig. \ref{fig:3}e and  \ref{fig:3}f refer to the case in which the basal membrane breaks due to the action of mechanical forces. Simulations of the rupture of the basal membrane induced by MMPs are reported in Fig. \ref{fig:mpp} and in
Movie S8. In particular, the Movie S8 illustrates how the effect of MMPs leads to a much rapid invasion of the dermis. This
is because MMPs typically break the basal membrane in different locations, while when the membrane ruptures mechanically
in one location the stress in the rest of the membrane is released preventing further rupture. Finally, we notice that the 
average compressive stress in the nevus is reduced when MMPs are present (see Fig. \ref{fig:mpp}b).

\subsubsection*{\it Stress dependent growth}
Inspired by the experimental results on primary melanoma that indicating that
compressive stress inhibits growth, we perform the simulations with a stress dependent
division rate, so that compressed cells divide less. The morphology of the resulting nevi changes only slightly with respect of the previous case (compare Fig. \ref{fig:6} with Fig.\ref{fig:3}), but we observe the development of incipient finger-like structures
(see in particular Fig.\ref{fig:6}c and \ref{fig:6}f).  We also measure the average compressive stress and find that
it is typically smaller than in the previous case (see Fig. \ref{fig:7}). This is due to the fact that in this case growth occurs in regions that
are less compressed so that there is more space to accommodate new cells.

\subsubsection*{\it Comparison with histological images}
\begin{figure}[ht]
\centering
\includegraphics[width=\textwidth]{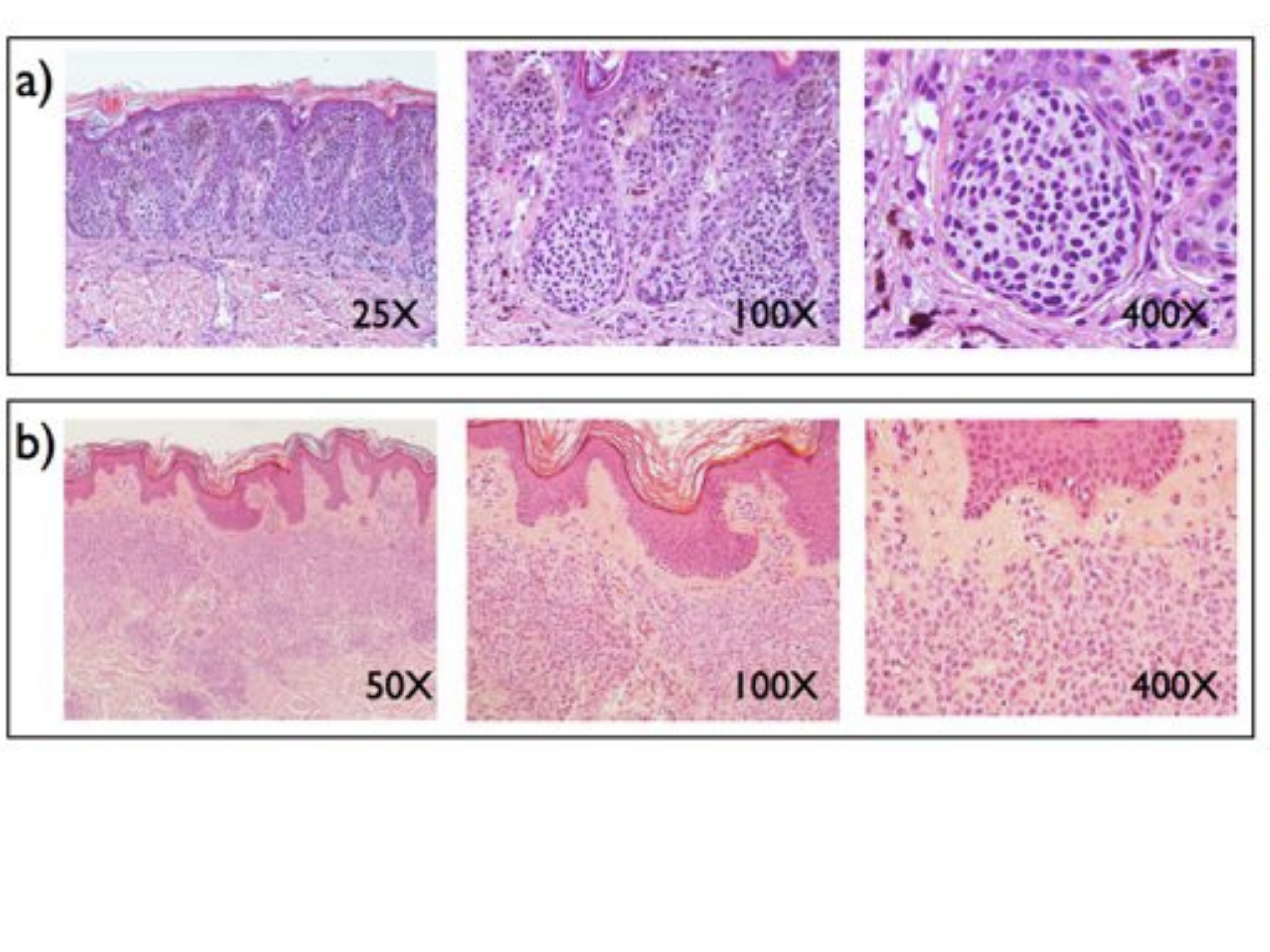}

\caption{{\bf Histological images of melanocytic nevi.} We illustrate the morphology observed by optical microscopy at different magnifications on histological sections of two types of nevi. a) Intraepithelial junctional nevi are confined in the epidermis and press against the basal membrane forming characteristic rounded pockets of densely packed cells (compare with Fig. \protect\ref{fig:3}b) b) Dermal nevi are confined in the dermis (compare with Fig. \protect\ref{fig:3}d) . Cells are loosely packed and do not touch the basal membrane. The basal membrane is located at the separation between the upper (epidermis) and lower (dermis) skin layers and indicated by an arrow in panel a).
Images courtesy of Dr. Claudio Clemente}
\label{fig:8}
\end{figure}

In Fig. \ref{fig:8}, we report two examples of common melanocytic nevi as revealed by histological sections.
In Fig. \ref{fig:8}a, we suggest that an interaepithelial juction nevus that is all confined in the epidermis
(compare with Fig. \ref{fig:3}b). Views at
different magnifications show that nevi cells are closely packed and press against the basal membrane as
indicated by rounded pocket shown in the left panel. In Fig. \ref{fig:8}b, we report an example of dermal nevus
(compare with Fig.  \ref{fig:3}d).
In this case, cells are less packed and are scattered all through the dermis without pressing on the basal membrane. These images are in good qualitative with the results of our model that show a tendency for nevi cells to press from the epidermis into the dermis rather than the reverse. This effect is likely due to the different mechanical properties of the two skin layers, as assumed in the model.

\section*{Discussion}

Melanoma is a very aggressive tumor, being chemio- and radio-resistant once it becomes metastatic  \cite{laporta2009b}. Its incidence is increasing both in Europe and in US, making it a current challenging problem for research. It is commonly believed  that nevi may progress into melanoma, or alternatively regress by differentiating \cite{clark1984}. Since diagnosis is mostly based on dermoscopic and histological analysis, it is extremely important to correlate the morphological features of nevi to the degree of malignancy or pre-malignancy of the lesions. Since nevi proliferate inside a complex microenvironment, we investigated the role of mechanics and geometry on the morphology and internal stresses of melanocytic nevi. Using human melanoma cell lines, we confirm also for melanoma a negative effect of the mechanical stress on cellular proliferation as already reported for other tumors \cite{helmlinger1997,cheng2009,montel2011,montel2012}.  We show that osmotic pressure is more effective in primary melanoma cells than in metastatic ones, suggesting that metastatic cells come from a subpopulation more aggressive and insensitive to the mechanical properties of the environment. We corroborate this result by experiments of cell growth in collagen coated plates where again primary cells are more affected than metastatic ones. This suggests that cells composing melanocytic nevi may be sensitive to stress as well, an issue that we investigated computationally.

We have devised a computational model for the growth of melanocytic nevi in a layered tissue representing the skin. Numerical simulations of the model show that the morphology of the resulting nevus depends considerably on the environment in which the growth takes place, in the epidermis close or far from the basal membrane or in the dermis. This is interesting since environmental properties, in this case the mechanical behavior of the tissue, is rarely considered in the progression of nevi. Simulations indicate that nevi are subject to fluctuating compressive stresses due to the tissue elasticity. If we introduce a stress dependence proliferation rate, we observe a decrease in the overall stress.  Our computational model is appropriate to describe a benign nevus, since we do not consider active motion of the cells but only their quasi-equilibrium displacement in response to elastic forces. To model the growth of melanoma cells, one should implement their mobility and  their ability to degrade the surrounding extracellular matrix, by expressing MMPs. We have incorporated
this last aspect in the simulations showing that the invasion of the dermis occurs more rapidly.  It is remarkable that many intriguing features observed  in melanoma, such as the formation of rough tumor boundaries and the invasion of the dermis from the epidermis, are observed from simple rules combining mechanics and geometry, although we simplified many of the irregularities present in the dermis. As a matter of fact we expect that, while the overall shape of the nevus can be influenced by the degree of accuracy in the dermis description, the qualitative picture remains unchanged.  A  limitation of our model lies in its two-dimensional nature that, while it greatly simplifies the numerical  calculations, could affect the values of the quantities we measure. We expect, however, that the general phenomenology we observe should be unchanged in a more realistic three-dimensional situation.

\section*{Materials and Methods}

\subsection*{\underline{Dextran solution}}
A master solution of dextran at 10\% (w/v) was formed  (Dextran from Leuconostoc spp, Fluka) and the diluted to the desired
concentration with complete medium. Transformation from dextran concentration to osmotic pressure was performed according to the calibration curve measured in Ref. \cite{bonnet-gonnet1994}.

\subsection*{\underline{Cell lines}}
Human IGR39 and IGR37 cells were obtained from Deutsche Sammlung von Mikroorganismen und Zellkulturen GmbH and cultured as previously described 
\cite{taghizadeh2010}. IGR39 was derived from a primary amelanotic cutaneous tumor and IGR37 from an inguinal lymph node metastasis in the same patient. 

\subsection*{\underline{Cell growth with collagen}}
 The cells were plated on pre-coated collagen type I dishes (Sigma) according to the manufacturer’s instructions.

\subsection*{\underline{Colony growth}}
 Cells are plated on 6 multiwells, fixed with 3.7\% paraformaldeide (PFA) for 5 minutes and then stained for 30min with 0.05\% crystal violet solution. After two washing with tap water, the plates are drained by inversion for a couple of minutes. In order to control the merging of different colonies, the experiments are performed with different initial cell concentrations as described recently by our group \cite{baraldi2013}. Data analysis of the resulting colonies has been performed according to Ref. \cite{baraldi2013}. 
 
\subsection*{\underline{Sulforhodamine B colorimetric assay for cytotoxicity screening}}
The sulforhodamine B (SRB) assay is used for cell density determination based on the cellular protein content according to Ref. \cite{vichai2006}

\subsection*{\underline{Apoptosis detection}}
The Annexin V-FITC Apoptosis Detection kit by Sigma was used to detect apoptotic and necrotic cells by detecting annexin V-FITC and propidium idodide by flow cytometry, respectively. Briefly the cells were incubated with Annexin V-FITC and propidium iodide  at room temperature for 10minutes and protect by light. Then the cells were immediately analysed by flow cytometry (FACSAria flow cytometer (Becton, Dickinson and Company, BD, Mountain View, CA). Data were analyzed using FlowJo software (Tree Star, Inc., San Carlos, CA).

\subsection*{\underline{Histological analysis of bioptic tissues}}
Tissue specimens are immediately fixed in neutral buffered formalin, embedded in paraffin,  sectioned and subjected to histopathological characterization after hematoxillin-eosin staining.

\subsection*{\underline{Statistical analysis}}
Statistical significance is evaluated according to the Kolmogorov-Smirnov non-parametric test with $p<10^{-2}$.

\subsection*{\underline{Model simulations}}
The model is simulated through a custom made python code.  To find mechanical equilibrium for the system, we use the fire relaxation method \cite{bitzek2006}. To speed up the code the relaxation routine has been written in C. Graphics rendering is done using Povray. Simulation and visualization codes are available at https://github.com/alexalemi/cancersim.

\section*{Acknowledgements:}
 We thank Fabien Montel for useful discussions and suggestions. We are specially grateful to Claudio Clemente for interesting discussions and for sharing with us his histological images.



\begin{thebibliography}{10}
\providecommand{\url}[1]{\texttt{#1}}
\providecommand{\urlprefix}{URL }
\expandafter\ifx\csname urlstyle\endcsname\relax
  \providecommand{\doi}[1]{doi:\discretionary{}{}{}#1}\else
  \providecommand{\doi}{doi:\discretionary{}{}{}\begingroup
  \urlstyle{rm}\Url}\fi
\providecommand{\bibAnnoteFile}[1]{%
  \IfFileExists{#1}{\begin{quotation}\noindent\textsc{Key:} #1\\
  \textsc{Annotation:}\ \input{#1}\end{quotation}}{}}
\providecommand{\bibAnnote}[2]{%
  \begin{quotation}\noindent\textsc{Key:} #1\\
  \textsc{Annotation:}\ #2\end{quotation}}
\providecommand{\eprint}[2][]{\url{#2}}

\bibitem{Baxter2013}
Baxter LL, Pavan WJ (2013) The etiology and molecular genetics of human
  pigmentation disorders.
\newblock Wiley Interdiscip Rev Dev Biol 2: 379-92.
\input{Baxter2013}

\bibitem{Kim2012}
Kim JK, Nelson KC (2012) Dermoscopic features of common nevi: a review.
\newblock G Ital Dermatol Venereol 147: 141-8.
\input{Kim2012}

\bibitem{Zalaudek2009}
Zalaudek I, Manzo M, Savarese I, Docimo G, Ferrara G, et~al. (2009) The
  morphologic universe of melanocytic nevi.
\newblock Semin Cutan Med Surg 28: 149-56.
\input{Zalaudek2009}

\bibitem{unna1893}
Unna PG (1893) Naevi and naevocarcinome.
\newblock Berl Klin Wochenschr 30.
\input{unna1893}

\bibitem{cramer1988}
Cramer SF (1988) The melanocytic differentiation pathway in congenital
  melanocytic nevi: theoretical considerations.
\newblock Pediatr Pathol 8: 253-65.
\input{cramer1988}

\bibitem{kittler2000}
Kittler H, Seltenheim M, Dawid M, Pehamberger H, Wolff K, et~al. (2000)
  Frequency and characteristics of enlarging common melanocytic nevi.
\newblock Arch Dermatol 136: 316-20.
\input{kittler2000}

\bibitem{weaver1997}
Weaver VM, Petersen OW, Wang F, Larabell CA, Briand P, et~al. (1997) Reversion
  of the malignant phenotype of human breast cells in three-dimensional culture
  and in vivo by integrin blocking antibodies.
\newblock J Cell Biol 137: 231-45.
\input{weaver1997}

\bibitem{helmlinger1997}
Helmlinger G, Netti PA, Lichtenbeld HC, Melder RJ, Jain RK (1997) Solid stress
  inhibits the growth of multicellular tumor spheroids.
\newblock Nat Biotech 15: 778--783.
\input{helmlinger1997}

\bibitem{cheng2009}
Cheng G, Tse J, Jain RK, Munn LL (2009) Micro-environmental mechanical stress
  controls tumor spheroid size and morphology by suppressing proliferation and
  inducing apoptosis in cancer cells.
\newblock PLoS ONE 4: e4632.
\input{cheng2009}

\bibitem{samuel2011}
Samuel MS, Lopez JI, McGhee EJ, Croft DR, Strachan D, et~al. (2011)
  Actomyosin-mediated cellular tension drives increased tissue stiffness and
  $\beta$-catenin activation to induce epidermal hyperplasia and tumor growth.
\newblock Cancer Cell 19: 776-91.
\input{samuel2011}

\bibitem{montel2011}
Montel F, Delarue M, Elgeti J, Malaquin L, Basan M, et~al. (2011) Stress clamp
  experiments on multicellular tumor spheroids.
\newblock Phys Rev Lett 107: 188102.
\input{montel2011}

\bibitem{montel2012}
Montel F, Delarue M, Elgeti J, Vignjevic D, Cappello G, et~al. (2012) Isotropic
  stress reduces cell proliferation in tumor spheroids.
\newblock New Journal of Physics 14: 055008.
\input{montel2012}

\bibitem{Tse2012}
Tse JM, Cheng G, Tyrrell JA, Wilcox-Adelman SA, Boucher Y, et~al. (2012)
  Mechanical compression drives cancer cells toward invasive phenotype.
\newblock Proc Natl Acad Sci U S A 109: 911-6.
\input{Tse2012}

\bibitem{paszek2005}
Paszek MJ, Zahir N, Johnson KR, Lakins JN, Rozenberg GI, et~al. (2005)
  Tensional homeostasis and the malignant phenotype.
\newblock Cancer Cell 8: 241-54.
\input{paszek2005}

\bibitem{hoffman2011}
Hoffman BD, Grashoff C, Schwartz MA (2011) Dynamic molecular processes mediate
  cellular mechanotransduction.
\newblock Nature 475: 316--323.
\input{hoffman2011}

\bibitem{zaman2006}
Zaman MH, Trapani LM, Sieminski AL, Siemeski A, Mackellar D, et~al. (2006)
  Migration of tumor cells in 3d matrices is governed by matrix stiffness along
  with cell-matrix adhesion and proteolysis.
\newblock Proc Natl Acad Sci U S A 103: 10889-94.
\input{zaman2006}

\bibitem{koch2012}
Koch TM, M{\"u}nster S, Bonakdar N, Butler JP, Fabry B (2012) 3d traction
  forces in cancer cell invasion.
\newblock PLoS One 7: e33476.
\input{koch2012}

\bibitem{Racz2007}
Racz B, Reglodi D, Fodor B, Gasz B, Lubics A, et~al. (2007) Hyperosmotic
  stress-induced apoptotic signaling pathways in chondrocytes.
\newblock Bone 40: 1536-43.
\input{Racz2007}

\bibitem{Nielsen2008}
Nielsen MB, Christensen ST, Hoffmann EK (2008) Effects of osmotic stress on the
  activity of {MAPK}s and {PDGFR}-beta-mediated signal transduction in {NIH}-{3T3}
  fibroblasts.
\newblock Am J Physiol Cell Physiol 294: C1046-55.
\input{Nielsen2008}

\bibitem{simonsen2012}
Simonsen TG, Gaustad JV, Leinaas MN, Rofstad EK (2012) High interstitial fluid
  pressure is associated with tumor-line specific vascular abnormalities in
  human melanoma xenografts.
\newblock PLoS One 7: e40006.
\input{simonsen2012}

\bibitem{wu2013}
Wu M, Frieboes HB, McDougall SR, Chaplain MAJ, Cristini V, et~al. (2013) The
  effect of interstitial pressure on tumor growth: coupling with the blood and
  lymphatic vascular systems.
\newblock J Theor Biol 320: 131-51.
\input{wu2013}

\bibitem{Walker2004}
Walker DC, Southgate J, Hill G, Holcombe M, Hose DR, et~al. (2004) The
  epitheliome: agent-based modelling of the social behaviour of cells.
\newblock Biosystems 76: 89-100.
\input{Walker2004}

\bibitem{Holcombe2012}
Holcombe M, Adra S, Bicak M, Chin S, Coakley S, et~al. (2012) Modelling complex
  biological systems using an agent-based approach.
\newblock Integr Biol (Camb) 4: 53-64.
\input{Holcombe2012}

\bibitem{galle2005}
Galle J, Loeffler M, Drasdo D (2005) Modeling the effect of deregulated
  proliferation and apoptosis on the growth dynamics of epithelial cell
  populations in vitro.
\newblock Biophysical Journal 88: 62 - 75.
\input{galle2005}

\bibitem{drasdo2005}
Drasdo D, H{\"o}hme S (2005) A single-cell-based model of tumor growth in vitro
  : monolayers and spheroids.
\newblock Physical Biology 2: 133.
\input{drasdo2005}

\bibitem{drasdo2012}
Drasdo D, Hoehme S (2012) Modeling the impact of granular embedding media, and
  pulling versus pushing cells on growing cell clones.
\newblock New Journal of Physics 14: 055025.
\input{drasdo2012}

\bibitem{Simpson2013}
Simpson MJ, Binder BJ, Haridas P, Wood BK, Treloar KK, et~al. (2013)
  Experimental and modelling investigation of monolayer development with
  clustering.
\newblock Bull Math Biol 75: 871-89.
\input{Simpson2013}

\bibitem{Treloar2013}
Treloar KK, Simpson MJ, Haridas P, Manton KJ, Leavesley DI, et~al. (2013)
  Multiple types of data are required to identify the mechanisms influencing
  the spatial expansion of melanoma cell colonies.
\newblock BMC Syst Biol 7: 137.
\input{Treloar2013}

\bibitem{Plank2012}
Plank MJ, Simpson MJ (2012) Models of collective cell behaviour with crowding
  effects: comparing lattice-based and lattice-free approaches.
\newblock J R Soc Interface 9: 2983-96.
\input{Plank2012}

\bibitem{chatelain2011}
Chatelain C, Balois T, Ciarletta P, Amar MB (2011) Emergence of microstructural
  patterns in skin cancer: a phase separation analysis in a binary mixture.
\newblock New Journal of Physics 13: 115013.
\input{chatelain2011}

\bibitem{Eikenberry2009}
Eikenberry S, Thalhauser C, Kuang Y (2009) Tumor-immune interaction, surgical
  treatment, and cancer recurrence in a mathematical model of melanoma.
\newblock PLoS Comput Biol 5: e1000362.
\input{Eikenberry2009}

\bibitem{Adra2010}
Adra S, Sun T, MacNeil S, Holcombe M, Smallwood R (2010) Development of a three
  dimensional multiscale computational model of the human epidermis.
\newblock PLoS One 5: e8511.
\input{Adra2010}

\bibitem{Thingnes2012}
Thingnes J, Lavelle TJ, Hovig E, Omholt SW (2012) Understanding the melanocyte
  distribution in human epidermis: an agent-based computational model approach.
\newblock PLoS One 7: e40377.
\input{Thingnes2012}

\bibitem{Hendriks_2003}
Hendriks FM, Brokken D, van Eemeren J, Oomens C,  Baajens F, et~al. (2003) A
  numerical-experimental method to characterize the non-linear mechanical
  behaviour of the human skin.
\newblock Skin Research and Technology 9: 274--283.
\input{Hendriks_2003}

\bibitem{Hendriks_phD_2005}
Hendriks FM (2005) Mechanical behaviour of human epidermal and dermal layers in
  vivo.
\newblock Technische Universiteit Eindhoven: Ph. D. Thesis.
\input{Hendriks_phD_2005}

\bibitem{Hendriks_2006}
Hendriks FM, Brokken D, Oomens C, Bader D, Baajens F (2006) The relative
  contributions of different skin layers to the mechanical behaviour of human
  skin in vivo using suction experiments.
\newblock Medical Engeneering \& Physics 28: 259--266.
\input{Hendriks_2006}

\bibitem{james1975}
James AG, Green A, Simpson G (1975) Strain energy functions of rubber. i.
  characterization of gum vulcanizates.
\newblock Journal of Applied Polymer Science 19: 2033--2058.
\input{james1975}

\bibitem{James_2000}
{MSC Software Corporation} (2001) Volume A: Theory and user information,
  Version 2001.
\newblock MSC.MARC, 1 edition.
\input{James_2000}

\bibitem{Long_2011}
Long R, Hall MS, Wu MM, Hui CH (2011) Effects of gel thickness on microscopic
  indentation measurements of gel modulus.
\newblock Biophysical Journal 101: 643--650.
\input{Long_2011}

\bibitem{lekka1999}
Lekka M, Laidler P, Gil D, Lekki J, Stachura Z, et~al. (1999) Elasticity of
  normal and cancerous human bladder cells studied by scanning force
  microscopy.
\newblock Eur Biophys J 28: 312-6.
\input{lekka1999}

\bibitem{drasdo1995}
Drasdo D, Kree R, McCaskill J (1995) Monte {C}arlo approach to tissue-cell
  populations.
\newblock Physical Review E 52: 6635-6657.
\input{drasdo1995}

\bibitem{baraldi2013}
Baraldi MM, Alemi AA, Sethna JP, Caracciolo S, La~Porta CAM, et~al. (2013)
  Growth and form of melanoma cell colonies.
\newblock Journal of Statistical Mechanics: Theory and Experiment 2013: P02032.
\input{baraldi2013}

\bibitem{howell1999}
Howell D, Behringer RP, Veje C (1999) Stress fluctuations in a 2d granular
  couette experiment: A continuous transition.
\newblock Phys Rev Lett 82: 5241--5244.
\input{howell1999}

\bibitem{laporta2009b}
La Porta CAM (2009) Mechanism of drug sensitivity and resistance in melanoma.
\newblock Curr Cancer Drug Targets 9: 391--397.
\input{laporta2009b}

\bibitem{clark1984}
Clark WH, Elder DE, Guerry D, Epstein MN, Greene MH, et~al. (1984) A
  study of tumor progression: the precursor lesions of superficial spreading
  and nodular melanoma.
\newblock Hum Pathol 15: 1147-65.
\input{clark1984}

\bibitem{bonnet-gonnet1994}
Bonnet-Gonnet C, Belloni L, Cabane B (1994) Osmotic pressure of latex
  dispersions.
\newblock Langmuir 10: 4012-4021.
\input{bonnet-gonnet1994}

\bibitem{taghizadeh2010}
Taghizadeh R, Noh M, Huh YH, Ciusani E, Sigalotti L, et~al. (2010) {CXCR6}, a
  newly defined biomarker of tissue-specific stem cell asymmetric self-renewal,
  identifies more aggressive human melanoma cancer stem cells.
\newblock PLoS One 5: e15183.
\input{taghizadeh2010}

\bibitem{vichai2006}
Vichai V, Kirtikara K (2006) Sulforhodamine {B} colorimetric assay for
  cytotoxicity screening.
\newblock Nat Protoc 1: 1112-6.
\input{vichai2006}

\bibitem{bitzek2006}
Bitzek E, Koskinen P, G\"ahler F, Moseler M, Gumbsch P (2006) Structural
  relaxation made simple.
\newblock Phys Rev Lett 97: 170201.
\input{bitzek2006}

\end{thebibliography}

\clearpage

\section*{Tables}
\begin{table}[h]
\begin{center}

\begin{tabular}{|c|c|c|c|c|}
\hline
&  \multicolumn{2}{|c|}{3 days} &
  \multicolumn{2}{c|}{6 days} \\
\hline
& Necrosis & Early Apoptosis & Necrosis & Early Apoptosis\\ 
\hline 
0 kPa & 0.47 & 1.17 & 0.47 & 9.48\\ 
\hline 
0.2 kPa & 0.36 & 0.32 & 0.38 & 6.51\\ 
\hline 
1 kPa & 1.1 & 0.94 & 1.91  & 9.56\\ 
\hline 
\end{tabular} 

\end{center}
\caption{{\bf Effect of osmotic pressure on cell death.} Percentage of IgR39 cells in necrosis or early apoptosis after 3 or 6 days for various osmotic
pressures in a typical experiment.}
\end{table}

\newpage
\section*{Figure captions}

\clearpage
\setcounter{figure}{0}
\renewcommand{\thefigure}{S\arabic{figure}}

\section*{Supplementary Information}

\subsection*{\underline{Supplementary figure captions}}

\begin{figure}[ht]
\centering
\includegraphics[width=\textwidth]{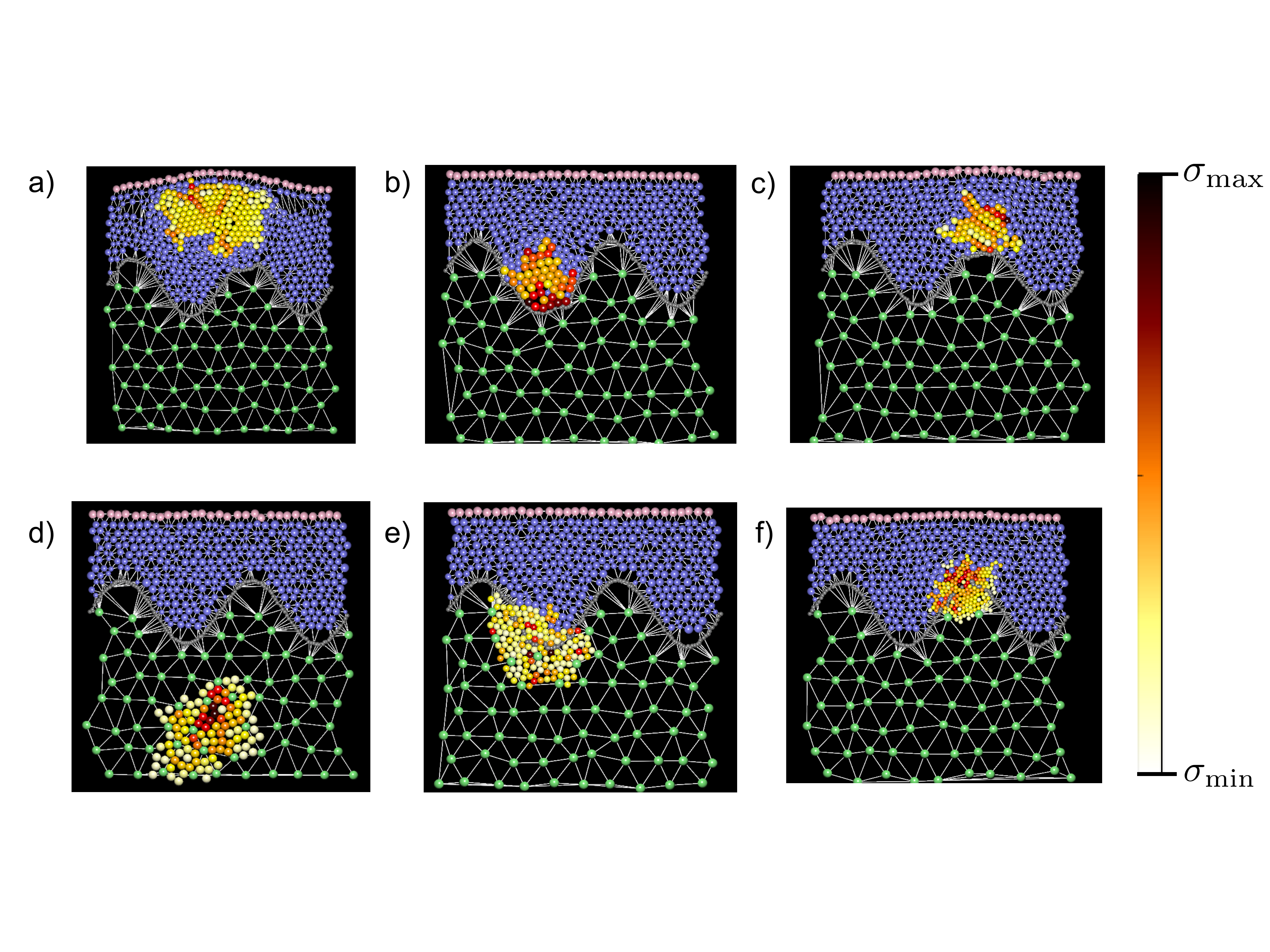}

\caption{{\bf Morphology of nevi for different locations of the initiating cell for pressure dependent growth.} Illustration of the results of numerical simulations for nevi grown  from melanocytes located in different positions in the skin and for different mechanical properties of the basal membrane. The conditions are the same as in Fig. \protect\ref{fig:3}, but here the growth depends on the compressive stress acting on each cell. 
Growing melanocytes are shown with a varying color that reflects their compressive stresses according to the color bar. The maximum $\sigma_{\rm max}$ and
minimum  $\sigma_{\rm min}$ for each configuration are equal to (in kPa):
a)7.46 $10^{-6}-0.13$, b)$8.24 10^{-6}-0.52$, c) $0.0016-0.63$ d)$0.00015-0.136$
e) $0.00039-0.043$
f) $0.0022-0.428$.
}
\label{fig:6}
\end{figure}

\begin{figure}[ht]
\centering
\includegraphics[width=\textwidth]{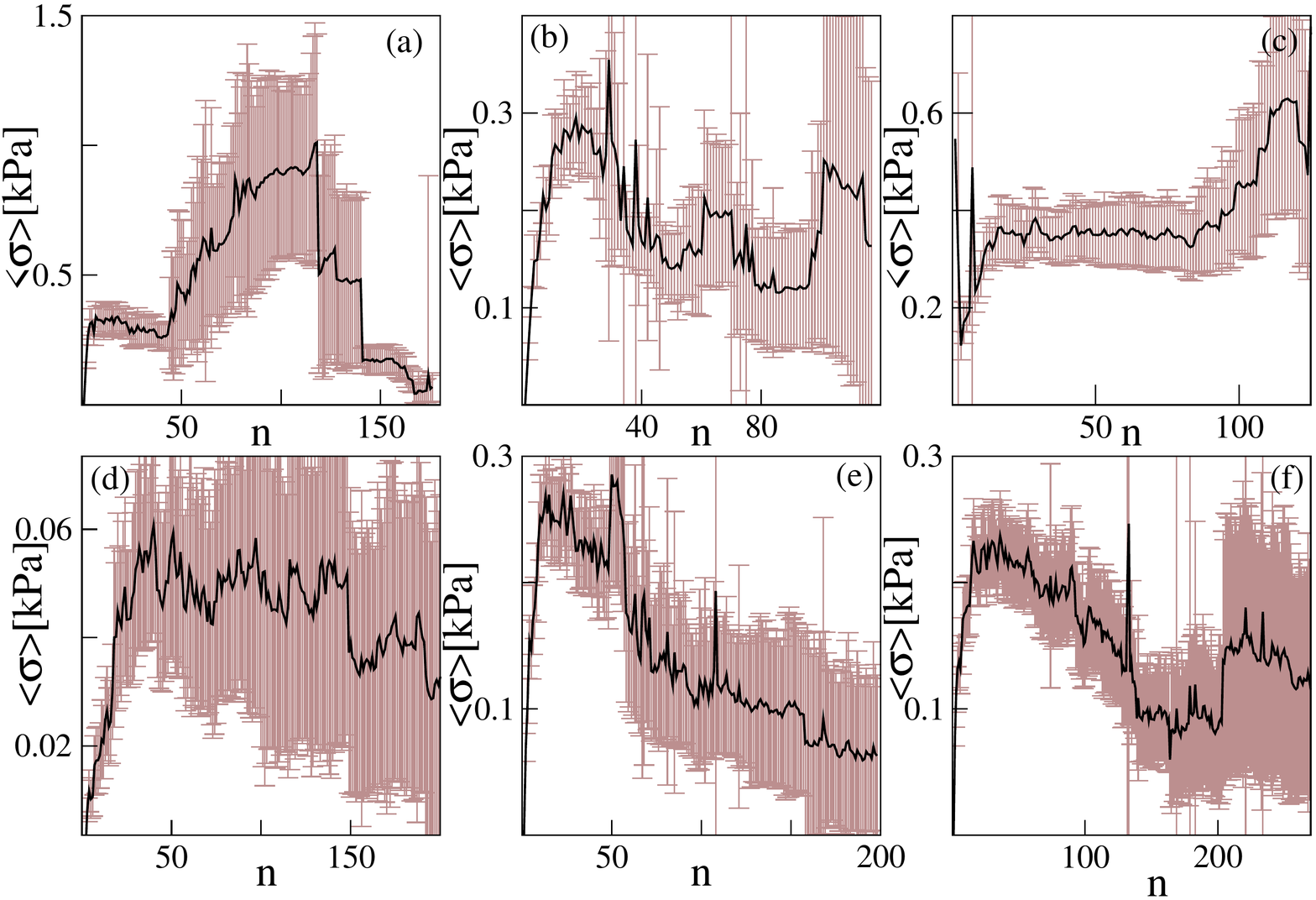}
\caption{{\bf Mechanical stresses in pressure dependent growth of nevi.} We report the evolution of the average compressive stress experienced by the melanocytes composing the nevus as a function of the size of the nevus, quantified
by the number of cells $n$,  for the same conditions as in Fig. \protect\ref{fig:6}. The error bars represent the 
standard error of the mean. The different panels represent different initial locations: a) in the middle of the epidermis, b) in the minima of a strong basal membrane, c) in the maxima of a strong basal membrane, d) in the dermis, e) in the minima of
a weak basal membrane, f) in the maxima of a weak basal membrane.}
\label{fig:7}
\end{figure}

\subsection*{\underline{Supplementary movie captions}}

{\bf Movie S1:} Growth of a nevus starting from a melanocyte placed deep inside the epidermis.

{\bf Movie S2:}  Growth of a nevus starting from a melanocyte placed in the epidermis at the minimum of a weak
basal membrane.

{\bf Movie S3:} Growth of a nevus starting from a melanocyte placed in the epidermis at the minimum of a strong
basal membrane.

{\bf Movie S4:} Growth of a nevus starting from a melanocyte placed in the epidermis at the maximum of a weak
basal membrane.

{\bf Movie S5:} Growth of a nevus starting from a melanocyte placed in the epidermis at the maximum of a strong
basal membrane.

{\bf Movie S6:} Growth of a nevus starting from a melanocyte placed deep inside the dermis.

{\bf Movie S7:} Growth of a nevus starting from a melanocyte placed deep inside the dermis considering a direct interaction
with the ECM.

{\bf Movie S8:} Growth of a nevus starting from a melanocyte placed in the epidermis at the maximum of a strong
basal membrane that can be broken by MMPs.

\end{document}